\documentclass[conference]{IEEEtran}

\usepackage{amsmath,amssymb}
\usepackage{booktabs}
\usepackage{graphicx}
\usepackage{url}
\usepackage[hidelinks]{hyperref}

\title{Training-Free Affinity Fusion of Neural and Embedding-Based Speaker Diarization}

\author{\IEEEauthorblockN{Yehoshua Dissen\textsuperscript{1,*}\
Joseph Keshet\textsuperscript{2}\,
Eduard Golshtein\textsuperscript{1,*}}
\IEEEauthorblockA{\textsuperscript{1}Linguana \qquad \textsuperscript{2}Technion -- Israel Institute of Technology\\
\textsuperscript{*}Equal contribution}}

\begin{document}
\maketitle

\begin{abstract}
Speaker diarization systems based on speaker embeddings and neural diarization exploit complementary forms of speaker information, but their intermediate representations are not directly compatible. We introduce Training-Free Affinity Fusion (TFAF), which integrates the speaker structure inferred by a neural diarizer into an embedding-based diarization system. The neural speaker partition is used to condition local speaker representations, from which we construct a continuous affinity matrix and combine it with the embedding-based acoustic affinity before a single global clustering step. The method requires no additional training, shared embedding space, speaker-label alignment, or hard transfer of the neural diarizer's speaker count. Experiments on AMI and CALLHOME show consistent DER improvements over both constituent systems; on AMI, fusion also improves speaker-attributed transcription. Ablations show that the neural speaker partition accounts for most of the gain, while retaining the continuous embedding-based affinities provides additional benefit over hard partition fusion.
\end{abstract}

\begin{IEEEkeywords}
speaker diarization, speaker clustering, affinity fusion, system combination
\end{IEEEkeywords}

\section{Introduction}Speaker diarization, the task of determining who spoke when, is commonly approached through two broad paradigms: clustering speaker embeddings over a recording, or using a neural diarization model to directly estimate speaker activity from the speech sequence~\cite{anguera2012review,park2022review}. Embedding-clustering systems are particularly well suited to maintaining speaker identity over long recordings. Their main limitation is the tradeoff between speaker discrimination and temporal resolution. Long analysis windows produce more reliable speaker embeddings but blur short turns and speaker changes, whereas short windows provide better temporal localization but weaker speaker representations. Multiscale clustering addresses this tradeoff by combining affinities computed at several window lengths before global spectral clustering~\cite{park2021multiscale,park2022msdd}.

Neural diarization systems make a different tradeoff. By processing sequences of speech jointly, they can use temporal context to resolve short turns, speaker transitions, and overlapping speech rather than treating each segment as an independent speaker observation. However, processing long recordings directly with a neural sequence model can require substantial memory and computation. Long recordings are therefore often processed in shorter chunks, which introduces the additional problem of maintaining consistent speaker identities across chunks. More generally, neural diarization must accommodate a variable number of speakers and the permutation ambiguity of speaker labels~\cite{fujita2019eend,kinoshita2021hybrid}. These properties make the two paradigms naturally complementary. Neural diarization provides detailed local speaker information, while embedding clustering provides strong recording-level speaker discrimination and global consistency~\cite{kinoshita2021hybrid}.

Combining the two paradigms is not straightforward because their intermediate speaker representations are generally incompatible. Embedding-based diarization operates on continuous pairwise speaker similarities, whereas neural diarization produces recording-internal speaker assignments and representations that are not assumed to share the embedding system's representation space. Consequently, the two systems cannot in general be fused by directly comparing their speaker labels or representations. Existing alternatives address related but different settings: DOVER and DOVER-Lap~\cite{stolcke2019dover,raj2021doverlap} fuse completed diarization outputs, co-association methods combine hard same-cluster relations~\cite{strehl2002cluster,yin2018dihard}, probability-level fusion requires compatible neural outputs~\cite{alvarez2025probabilistic}, and hybrid neural--clustering systems integrate the two paradigms within a single diarization pipeline~\cite{kinoshita2021hybrid}. What is missing is a training-free interface that transfers neural speaker structure into an embedding-based affinity graph while preserving its continuous acoustic information.

We introduce Training-Free Affinity Fusion (TFAF), which uses the neural speaker partition to construct a continuous, partition-conditioned affinity that interfaces with the embedding-based graph. The neural diarizer's recording-internal speaker structure conditions its local representations to form soft affinity evidence, which is added to the continuous embedding-based affinity before final clustering. This preserves the acoustic geometry of the embedding system while allowing the neural diarizer to contribute speaker information without label alignment, a shared embedding space, additional training, or hard adoption of its speaker count or final partition. Experiments on AMI and CALLHOME show that TFAF improves over both systems individually. Ablations show that most of the transferable information comes from the neural speaker partition, while preserving the continuous acoustic affinity is important for effective fusion.

\section{Related Work}

A wide range of approaches have been explored for embedding-based diarization, from x-vector speaker representations~\cite{snyder2018xvectors} to stronger neural encoders such as TitaNet~\cite{koluguri2021titanet}, multiscale methods that combine information across different temporal resolutions~\cite{park2021multiscale,park2022msdd}, and self-supervised approaches that learn speaker representations from unlabeled speech~\cite{dissen2022selfsupervised,dissen2026label}.

Neural diarization instead estimates speaker activity directly from sequences of speech~\cite{fujita2019eend}. By processing temporal context jointly, these systems can model short turns, speaker transitions, and overlapping speech without treating each segment as an independent speaker observation. Hybrid approaches combine neural local diarization with clustering or speaker tracking to maintain speaker identities over longer recordings and accommodate a variable number of speakers~\cite{kinoshita2021hybrid,han2025diarizen}.

Several works combine complementary information at different stages of the diarization pipeline. Park et al.~\cite{park2019lexical} combine lexical and acoustic adjacency matrices before spectral clustering, while Turn-to-Diarize~\cite{xia2022turn} uses detected speaker turns to constrain embedding-based diarization. At the system-output level, DOVER and DOVER-Lap align speaker labels across completed diarization hypotheses and combine their decisions~\cite{stolcke2019dover,raj2021doverlap}. Co-association methods instead convert hard partitions into pairwise same-cluster evidence~\cite{strehl2002cluster,yin2018dihard}, while recent probability-level fusion combines calibrated speaker-activity probabilities from multiple neural diarization systems~\cite{alvarez2025probabilistic}.

Our approach differs from these methods in both the information it combines and the stage at which fusion is performed. The embedding-based system provides a continuous acoustic affinity matrix, while the neural diarizer provides speaker assignments and local neural representations which are not assumed to share a representation space with the embedding system. Rather than aligning completed outputs or converting both systems into hard partitions, we preserve the continuous acoustic affinities and incorporate the neural speaker structure before final clustering. This avoids speaker-label alignment and hard transfer of the neural diarizer's speaker count or final partition.

\section{Method}

Embedding-based and neural diarization systems provide complementary speaker evidence. Embedding-based systems compare speaker embeddings over the recording, providing strong global speaker discrimination, but they inherit the tradeoff between short windows for temporal resolution and longer windows for reliable speaker embeddings. One way to overcome this, is by using multiple temporal scales and combine them.  Neural diarizers use temporal context to model speaker activity and speaker turns more directly, but expose recording-internal speaker assignments rather than the same continuous affinity structure used by embedding clustering.

Our approach combines these two sources at the affinity level before the final clustering step. Let the embedding-based system define $N$ base segments. At acoustic scale $s$, speaker embeddings are extracted from scale-specific windows anchored to the same base segments, producing $A^{(s)}\in\mathbb{R}^{N\times N}$. The embedding-based affinity is
\begin{equation}
A^{\mathrm{emb}} = \sum_{s=1}^{K} w_s A^{(s)},
\label{eq:embedding_affinity}
\end{equation}
where $K$ is the number of acoustic scales. We use equal weights $w_s=1$, so $A^{\mathrm{emb}}$ is the sum of the per-scale affinities.

The neural diarizer supplies local speaker representations $\ell_q$, a recording-level assignment $k(q)$ for each local speaker instance $q$, and frame-level speaker activity indicating when each $q$ is active. The labels $k(q)$ are recording-internal and carry no correspondence to the clusters produced by the embedding-based system. We do not require pairwise similarity between the local representations to define a globally consistent speaker metric; when it is informative, the fusion exploits it, while the partition-only variant below requires only the speaker assignments.

For each inferred neural speaker $m$, let $c_m$ be the centroid of the local representations assigned to that speaker. Before blending, both vectors are normalized, $\hat{\ell}_q=\ell_q/\lVert\ell_q\rVert_2$ and $\hat c_m=c_m/\lVert c_m\rVert_2$. We form the partition-conditioned representation
\begin{equation}
u_q =
\frac{(1-\alpha)\hat{\ell}_q + \alpha \hat c_{k(q)}}
{\left\lVert(1-\alpha)\hat{\ell}_q + \alpha \hat c_{k(q)}\right\rVert_2}.
\label{eq:partition_conditioned_rep}
\end{equation}
Thus $\alpha=0.5$ gives the local and speaker-level representations equal vector weight.

The $u_q$ representations are mapped onto a fixed neural temporal grid by accumulating their contributions over the 20 ms frames in which the corresponding local speaker is active.
Frames with exactly one active local speaker contribute its $u_q$; frames with no active speaker or multiple active speakers contribute no evidence. The contributions within each neural cell are averaged and normalized to obtain $z_i$ on the base-segment grid. Empty cells are set to $z_i=0$, so the neural branch abstains rather than contributing different-speaker evidence. A segment crossing a neural speaker boundary can therefore contain contributions from both speakers rather than receiving a hard label.

We construct a continuous neural affinity matrix $B$, with $B_{ij}$ equal to the cosine similarity between nonzero $z_i$ and $z_j$ and zero if either segment is uncovered. We fuse the two affinity sources as
\begin{equation}
A^{\mathrm{fused}}
= A^{\mathrm{emb}} + \lambda B
= \sum_{s=1}^{K} w_s A^{(s)} + \lambda B,
\label{eq:fused_affinity}
\end{equation}
where $\lambda$ controls the neural contribution. The same clustering procedure used by the embedding-based system is then applied once to $A^{\mathrm{fused}}$ to obtain the final speaker partition and speaker count. Meaning, the neural diarizer contributes speaker evidence without imposing its labels or speaker count on the final output.

For the partition-only ablation, $B$ is replaced by a same-speaker affinity derived from the neural frame assignments. The frame-level indicator is $\mathbf{1}[k(q(t))=k(q(t'))]$ when both frames have a single active speaker and zero otherwise, and is aggregated using the same temporal mapping. Boundary cells may therefore yield fractional affinities. In the full system, $\alpha=0.5$ and $\lambda=1$ unless otherwise stated.

\section{Experimental Setup}

\subsection{Systems}

Our embedding-based system is the NeMo Equal-w-MS-Clus configuration~\cite{park2022msdd}, using TitaNet-L speaker embeddings~\cite{koluguri2021titanet} and NME-SC~\cite{park2020nme,kuchaiev2019nemo}. For AMI, we use six window durations, $[3.0, 2.5, 2.0, 1.5, 1.0, 0.5]$~s, with half-window shifts. For CALLHOME, we use the corresponding five telephony durations, $[1.5, 1.25, 1.0, 0.75, 0.5]$~s~\cite{park2022msdd}. VBx~\cite{landini2022vbx} is used only as the additional hypothesis in the three-system fusion baselines.

Our neural diarization system is DiariZen~\cite{han2025diarizen}. We use the frozen public \texttt{diarizen-wavlm-large-s80-md} checkpoint with its default pipeline and speech detection. DiariZen processes the recording in 16~s chunks with a 1.6~s step. Its segmentation model produces speaker activities every 20~ms for up to four chunk-local speakers. For each chunk-local speaker, the speaker-embedding stage extracts a 256-dimensional representation $\ell_q$ from the corresponding activity-masked audio, and DiariZen's clustering stage assigns that local speaker to a recording-level speaker $k(q)$. For each recording-level speaker $m$, we compute the centroid $c_m$ from the local representations assigned to that speaker.

To combine the DiariZen output with Equal-w-MS-Clus, we map the neural information onto a 0.5~s window / 0.25~s shift grid. Each 20~ms frame with exactly one active DiariZen speaker contributes the corresponding partition-conditioned representation $u_q$ from Eq.~\eqref{eq:partition_conditioned_rep} to every grid cell containing that frame. Each cell is represented by the $\ell_2$-normalized mean of its contributions; cells with no single-speaker frames are assigned the zero vector and therefore contribute no neural affinity. Frames with multiple active speakers are excluded.

The resulting grid is treated as an additional affinity scale. The standard Equal-w-MS-Clus midpoint mapping associates each base-scale segment with the nearest neural grid cell, using the same temporal mapping employed for the acoustic scales. Consequently, a segment spanning a DiariZen speaker change receives a normalized mixture of the corresponding speaker representations rather than a hard speaker label. No additional training or learned alignment is introduced. 

\subsection{Datasets and scoring}

On AMI~\cite{carletta2006ami}, we use the official 16-meeting MixHeadset test split and the only-words reference setup. Following the matched Equal-w-MS-Clus evaluation, we use oracle speech activity, ignore overlap, and score DER with a 0.25s collar. On CALLHOME~\cite{canavan1997callhome}, we use SRE-2000 disc 8 with the matched telephony configuration and oracle speech activity for the main comparisons. DiariZen itself is always run with its default speech detection; for oracle-SAD comparisons, all system outputs are restricted to the same oracle speech regions before scoring.

For speaker-attributed evaluation, the recognized transcript is held fixed across diarization systems. \emph{Word err.} counts reference words assigned to the wrong speaker after optimal speaker-label mapping, excluding overlap-ambiguous words. We also report time-constrained minimum-permutation WER (tcpWER) with a 5s collar using MeetEval~\cite{vonNeumann2023meeteval}; unlike Word err., tcpWER includes both recognition and speaker-attribution errors. For oracle-SAD experiments, systems that do not natively use oracle speech regions are adapted to the same speech coverage before scoring.

\section{Results}

We first evaluate whether TFAF improves over its two constituent systems. On AMI, our Equal-w-MS-Clus baseline closely matches the published result of Park et al.~\cite{park2022msdd} (1.08 vs.\ 1.06 DER). On CALLHOME, our implementation obtains 3.85 DER, compared with 4.57 reported in~\cite{park2022msdd}. As shown in Table~\ref{tab:cross_corpus}, TFAF improves over both systems on both corpora.

\begin{table}[t]
\caption{Cross-corpus performance under identical oracle speech coverage.}
\label{tab:cross_corpus}
\centering
\resizebox{\columnwidth}{!}{%
\begin{tabular}{llrr}
\toprule
Corpus & System & DER $\downarrow$ & Word err. $\downarrow$ \\
\midrule
AMI & Equal-w-MS-Clus & 1.08 & 739 \\
AMI & DiariZen & 1.54 & 859 \\
AMI & TFAF & \textbf{0.85} & \textbf{465} \\
\midrule
CALLHOME & Equal-w-MS-Clus & 3.85 & -- \\
CALLHOME & DiariZen & 4.17 & -- \\
CALLHOME & TFAF & \textbf{2.74} & -- \\
\bottomrule
\end{tabular}%
}
\end{table}

We next examine which part of the DiariZen output drives this improvement. Table~\ref{tab:information_ladder} uses the same DiariZen inference throughout and changes only the information transferred to Equal-w-MS-Clus. Transferring its estimated speaker count as a hard constraint performs poorly, increasing AMI Word err.\ from 739 to 3598. In contrast, the local neural representations already improve over the baseline, and introducing the speaker partition reduces the error further. The binary partition alone reaches 521 errors, accounting for most of the improvement, while adding speaker-centroid information gives the best result of 465 with TFAF.

\begin{table}[t]
\caption{AMI ablation under identical oracle speech coverage. Each ``+ DZ'' row adds the indicated information from the same DiariZen inference.}
\label{tab:information_ladder}
\centering
\resizebox{\columnwidth}{!}{%
\begin{tabular}{lrr}
\toprule
Setting & Word err. $\downarrow$ & tcpWER $\downarrow$ \\
\midrule
Equal-w-MS-Clus & 739 & 30.88 \\
DiariZen & 859 & 31.62 \\
+ DZ speaker count & 3598 & 37.03 \\
+ DZ local affinity & 531 & 30.25 \\
+ DZ binary partition affinity & 521 & \textbf{29.92} \\
+ DZ centroid affinity & 478 & 29.93 \\
+ DZ local+centroid affinity (TFAF) & \textbf{465} & 29.95 \\
\bottomrule
\end{tabular}%
}
\end{table}

CALLHOME shows the same pattern: the binary partition reduces DER from 3.85 to 2.90 and TFAF reaches 2.74 (Table~\ref{tab:callhome_ladder}). DiariZen estimates the exact CALLHOME speaker count on 83.6\% of files, compared with 71.7\% for Equal-w-MS-Clus, yet imposing that count raises DER to 4.85. The failure therefore comes from using the estimate as a hard global constraint rather than from poor count accuracy. The fusion weight is also not sharply tuned: $\lambda=0.5,1,$ and $2$ give DERs of 3.04, 2.74, and 2.69, respectively, so the fixed value $\lambda=1$ remains within 0.05 DER points of the best tested value.

\begin{table}[t]
\caption{CALLHOME ablation under identical oracle speech coverage.}
\label{tab:callhome_ladder}
\centering
\begin{tabular}{lr}
\toprule
Setting & DER $\downarrow$ \\
\midrule
Equal-w-MS-Clus & 3.85 \\
DiariZen & 4.17 \\
+ DZ speaker count & 4.85 \\
+ DZ binary partition affinity & 2.90 \\
+ DZ local+centroid affinity (TFAF) & \textbf{2.74} \\
\bottomrule
\end{tabular}
\end{table}

To test whether the systems provide complementary information, we condition on the Equal-w-MS-Clus acoustic affinity and ask whether DiariZen still predicts the true speaker relation. For pairs with acoustic affinity between 0.20 and 0.25, pairs assigned to the same DiariZen speaker belong to the same reference speaker with probability 0.902, compared with only 0.017 when DiariZen assigns them to different speakers. This separation remains at least $4.6\times$ for acoustic affinities below 0.40. The complementarity also works in the opposite direction: among pairs that DiariZen assigns to different speakers, the acoustic affinity distinguishes false splits (true same-speaker pairs) from true different-speaker pairs with an AUC of 0.858. Thus, each system provides useful information where the other is uncertain or incorrect.

We next ask which errors in the transferred partition are most harmful. We synthetically split DiariZen speaker clusters and measure how the resulting partition affects the final diarization. We summarize the severity of a split by $r_{\mathrm{cut}}=N_{\mathrm{cut}}/N_{\mathrm{same}}$, the fraction of true same-speaker pairwise relations removed by the corrupted partition. Across 432 synthetic splits, the increase in attribution error is strongly correlated with $r_{\mathrm{cut}}$ (Spearman $\rho=0.80$). Moreover, for similar values of $r_{\mathrm{cut}}$, coherent balanced splits are more damaging than small fragments because they create a plausible competing speaker partition. False merges are less harmful in our experiments, since the incorrect relations they add must compete with the existing acoustic structure rather than cutting through a true speaker cluster.

We then compare these controlled failures with DiariZen's actual errors. Its pooled $r_{\mathrm{cut}}$ is only 0.018, most of its fragmentation consists of small incoherent pieces, and its false-merge rate is 0.77\%. DiariZen therefore tends to make errors in the regime that our corruption experiment finds least damaging, which helps explain why its partition can improve the acoustic clustering despite being imperfect.

We next compare TFAF with alternative fusion strategies. Co-association~\cite{strehl2002cluster,yin2018dihard} first converts completed diarization outputs into hard same-speaker partitions and then reclusters them, while DOVER-Lap~\cite{raj2021doverlap} combines completed diarization hypotheses after label alignment. TFAF instead retains the continuous Equal-w-MS-Clus acoustic graph and incorporates the DiariZen information before the final global clustering decision.

\begin{table}[t]
\caption{Comparison of fusion levels under identical oracle speech coverage. Two-system co-association combines Equal-w-MS-Clus and DiariZen; three-system methods additionally use VBx.}
\label{tab:fusion_levels}
\centering
\resizebox{\columnwidth}{!}{%
\begin{tabular}{lrrrrr}
\toprule
System & DER$_{.25}$ & DER$_{0}$ & Word err. & tcpWER & CH DER \\
\midrule
Equal-w-MS-Clus & 1.08 & 2.35 & 739 & 30.88 & 3.85 \\
Co-association (2) & 1.31 & 3.02 & 889 & 31.61 & 3.97 \\
Co-association (3) & 1.24 & 2.90 & 701 & 30.68 & 4.35 \\
DOVER-Lap (3 systems) & \textbf{0.81} & 2.06 & 665 & 30.46 & 3.18 \\
TFAF & 0.85 & \textbf{1.95} & \textbf{465} & \textbf{29.95} & \textbf{2.74} \\
\bottomrule
\end{tabular}%
}
\end{table}

The co-association results show that pairwise encoding by itself is not sufficient. In particular, the two-system version first replaces the continuous Equal-w-MS-Clus graph with its hard partition and is worse than the unfused baseline on every reported AMI metric. Adding VBx improves this consensus, but it remains worse than affinity-level fusion. The important distinction is therefore not simply whether speaker relations are represented pairwise, but whether the continuous acoustic geometry of the primary system is preserved.

We also evaluated DOVER-Lap with the same two input systems, Equal-w-MS-Clus and DiariZen, where it gives no measurable improvement over Equal-w-MS-Clus on AMI. Because DOVER-Lap is more meaningful with multiple hypotheses, Table~\ref{tab:fusion_levels} reports the stronger three-system configuration with an additional VBx hypothesis, which improves over Equal-w-MS-Clus on both corpora.

On AMI, three-system DOVER-Lap obtains a slightly lower collared DER than TFAF (0.81 versus 0.85). TFAF instead gives substantially fewer wrong-speaker words (465 versus 665) and lower tcpWER (29.95 versus 30.46). Removing the DER collar also reverses the ordering, from 0.81 versus 0.85 to 2.06 versus 1.95. On CALLHOME, TFAF obtains 2.74 DER compared with 3.18 for DOVER-Lap.

The difference between collared DER and the transcription metrics is concentrated near speaker boundaries. Approximately two thirds of the attribution errors occur within 250~ms of a reference speaker boundary, precisely the region excluded by the standard DER collar. Small changes in collared DER therefore need not reflect changes in speaker-attributed transcription quality.

\section{Discussion and Conclusion}

We presented TFAF, a training-free interface for combining neural and embedding-based diarization systems whose speaker labels and representation spaces are otherwise incompatible. TFAF transfers recording-level neural speaker structure as soft affinity evidence while preserving the continuous acoustic graph used for global clustering. It improves over both constituent systems on AMI and CALLHOME, and ablations show that most of the transferable information comes from the neural speaker partition. Future work will extend TFAF to overlapping speech.

\end{document}